\documentclass[12pt]{article}
\usepackage{amscd,amsmath,amssymb,amstext,amsthm,exscale,latexsym}
\usepackage{graphicx}
\newcommand {\al}   {\alpha}       \newcommand {\bt}  {\beta}
\newcommand {\g }   {\gamma}       
\newcommand {\dl}   {\delta}       \newcommand {\e }  {\epsilon}
\newcommand {\z }   {\zeta}        \newcommand {\et}  {\eta}
\newcommand {\ve}   {\varepsilon}  
\newcommand {\lm}   {\lambda}

\newcommand {\vf }  {\varphi}      
         \newcommand {\om}  {\omega}
\newcommand {\Lm}   {\Lambda}      
       
\newcommand {\pl}   {\partial}     \newcommand {\nb}  {\nabla}
\newcommand {\wn}   {\widetilde\nabla}
\renewcommand {\div}{{\sf\,div\,}}       \renewcommand {\lim}{{\sf\,lim\,}}

\newcommand   {\const}{{\sf\,const}}     \newcommand   {\diag}{{\sf\,diag\,}}

\newcommand {\MM}  {{\mathbb M}}   
\newcommand {\MO}  {{\mathbb O}}   
   
\newcommand {\MS}  {{\mathbb S}}

\newcommand {\Sk}  {{\textsc{k}}}

\newcommand {\vol}  {\sqrt{|g|}}
\theoremstyle{definition}

\begin{document}
\title     {Gravity with dynamical torsion}
\author    {M.~O.~Katanaev\thanks{E-mail: katanaev@mi-ras.ru}
            \\ \sl Steklov Mathematical Institute of Russian Academy of
            Sciences,\\
            \sl 8 Gubkina St., Moscow 119991, Russia}
\maketitle
\begin{abstract}
We propose four simple Lagrangians for gravity models with dynamical torsion
which are free from ghosts and tachyons. The torsion propagates two massive or
massless particles of spin $1^\pm$ and $0^\pm$ besides the massless graviton
$2^+$ propagated by metric.
\end{abstract}
%******************************************************************************
\section{Introduction}
%*******************************************************************************
Introduction of dynamical torsion in general relativity is the natural geometric
generaliza\-tion of Einstein's gravity. It is the subject of research for many
years. Physical motivations and earlier references can be found in
\cite{HeHeKeNe76,HeMcMiNe95,Shapir02,Blagoj02,CapLau11}. In spite of much
efforts, we still do not have a generally acknowledged action for such model.
It is usually assumed that gravity
interactions are described by metric propagating spin $2^+$ massless gravitons,
whereas torsion describes massive or massless particles of spin $2^\pm$,
$1^\pm$, or $0^\pm$. To answer the crucial question whether torsion can yield a
physically acceptable generalization, we need a model but it is missing,
rather we have too many proposals and cannot make a unique choice. To my mind,
the main problem here is the mathematical complexity.

Torsion has three tensor indices and can propagate several massive or massless
particles of spin not higher then 2. From physical standpoint, the model must
not include ghosts and tachyons. A number of such models were proposed forty
years ago \cite{Nevill78,SezNie80,Sezgin81,HayShi80D}, but they are not analyzed
until now if full detail, mainly, because of technical difficulties. Moreover,
these models were found within a certain class of Lagrangians which do not
exhaust all possibilities. In the present paper, we consider another class of
dynamical torsion models without higher derivatives, considering metric and
torsion as independent variables, and found four Lagrangians without ghosts and
tachyons. The final answer seems to be much simpler and transparent then
previous proposals.
%******************************************************************************
\section{General Lagrangian}
%*******************************************************************************
We consider a space-time manifold $\MM$ with local coordinates $x^\al$,
$\al=0,1,2,3$, and the Riemann--Cartan geometry. The geometry can be described
either by metric $g_{\al\bt}$ and torsion $T_{\al\bt}{}^\g=-T_{\bt\al}{}^\g$ or,
equivalently, by Cartan variables: vierbein $e_\al{}^a$ and Lorentz connection
$\om_\al{}^{ab}=-\om_\al{}^{ba}$, $a,b=0,1,2,3$.

As usual, a vierbein defines unique metric
\begin{equation*}
  g_{\al\bt}:=e_\al{}^ae_\bt{}^b\eta_{ab},\qquad\eta_{ab}:=\diag(+---),
\end{equation*}
and the Lorentz connection is defined through the affine connection
$\Gamma_{\al\bt}{}^\g$ by the vielbein postulate
\begin{equation}                                                  \label{abxjhs}
  \nb_\al e_\bt{}^a=\pl_\al e_\bt{}^a-\Gamma_{\al\bt}{}^\g e_\g{}^a
  +e_\bt{}^b\om_{\al b}{}^a=0.
\end{equation}
If the vierbein is given, then this equation provides the one-to-one
correspondence between Lorentz connection and the torsion
$T_{\al\bt}{}^\g:=\Gamma_{\al\bt}{}^\g-\Gamma_{\bt\al}{}^\g$.

The vierbein has 6 components more then the metric due to the extra symmetry
with respect to local Lorentz transformations acting on Latin indices
$a,b,\dotsc$. In what follows, transformation of Latin indices into Greek ones
and vice versa is always performed by the vierbein and its inverse $e^\al{}_a$.
Moreover we assume that coordinates are chosen in such a way that the coordinate
$x^0$ is timelike and all sections $x^0=\const$ are spacelike.

Torsion and curvature have the following expressions in terms of Cartan
variables
\begin{align}                                                     \label{ecurct}
  T_{\al\bt}{}^a&=\pl_\al e_\bt{}^a-e_\al{}^b\om_{\bt b}{}^a
                    -(\al\leftrightarrow\bt),
\\                                                                \label{ecucac}
  R_{\al\bt a}{}^b&=\pl_\al \om_{\bt a}{}^b-\om_{\al a}{}^c
                       \om_{\bt c}{}^b-(\al\leftrightarrow\bt).
\end{align}

If we want vierbein and Lorentz connection to be really independent propagating
fields then they must satisfy differential equations of motion. Let us require
the field equations to be covariant and of second order. Then the Lagrangian
should be quadratic in curvature and torsion. The general form of such
Lagrangian is
\begin{align}                                                          \nonumber
  \frac1\vol L=&\kappa R-\frac14T^{abc}\big(\bt_1T_{abc}+\bt_2T_{cab}
  +\bt_3\eta_{ac}T_b\big)-
\\                                                                \label{anbcnd}
  &-\frac14R^{abcd}\big(\g_1R_{abcd}+\g_2R_{cdab}+\g_3R_{acbd}
  +\g_4\eta_{bd}R_{ac}+\g_5\eta_{bd}R_{ca}\big)-2\Lm,
\end{align}
where $T_b:=T_{ab}{}^b$ is the trace of torsion, $R_{ab}:=R_{acb}{}^c$ and
$R:=R_a{}^a$ are the Ricci tensor and scalar curvature, respectively; $\kappa$,
$\bt_{1,2,3}$, and $\g_{1,2,3,4,5}$ are coupling constants, and $\Lm$ is a
cosmological constant.
We do not have to add the scalar curvature $\widetilde R(e)$ built from
Christoffel's symbols for zero torsion due to the identity
\begin{equation}                                                  \label{eident}
  R+\frac14T_{\al\bt\g}T^{\al\bt\g}-\frac12T_{\al\bt\g}T^{\g\al\bt}-T_\al T^\al
    -\frac2\vol\pl_\al\big(\vol T^\al\big)=\widetilde R.
\end{equation}
Here and in what follows the tilde sign marks geometric notions constructed
entirely from metric for zero torsion.

There are three independent invariants quadratic in torsion in Lagrangian
(\ref{anbcnd}) because torsion has three irreducible components with respect to
the Lorentz group. Curvature tensor in Riemann--Cartan geometry has six
irreducible compo\-nents but we added only five in Eq.(\ref{anbcnd}) due to the
Gauss--Bonnet type identity
\begin{equation}                                                  \label{ancbdy}
  R_{abcd}R^{cdab}-4R_{ab}R^{ba}+R^2\equiv\widetilde\nb_\al X^\al,
\end{equation}
where $X^\al$ are components of some vector.

Several models without ghosts and tachyons were found within the class of
Lagrangians (\ref{anbcnd}) in \cite{SezNie80,Sezgin81,HayShi80D}, and the
analysis is rather complicated. There is the additional problem. Lagrangian
(\ref{anbcnd}) is not the most general one in  dynamical torsion models, because
the existence or absence of higher derivatives depend on the chosen independent
variables. For example, transformation of the Lorentz connection into torsion
tensor by Eq.(\ref{abxjhs}) involves derivatives of the
vierbein. Therefore Lagrangian (\ref{anbcnd}) is a higher derivative one for the
vierbein-torsion variables.

In what follows, we need the decomposition of torsion tensor $T_{abc}$ into
irreducible pieces with respect to the Lorentz group. To this end we extract the
trace $T_a$ and totally antisymmetric part $T_{[abc]}$:
\begin{equation}                                                  \label{etorci}
  T_{abc}=S_{abc}+\frac13(\et_{ac}T_b-\et_{bc}T_a)+T_{[abc]},
\end{equation}
where
\begin{equation*}
  T_b:=T_{ab}{}^a,\qquad T_{[abc]}:=\frac13(T_{abc}+T_{cab}+T_{bca}).
\end{equation*}
The irreducible component $S_{abc}$ is defined by the algebraic relations:
\begin{equation}                                                  \label{ehdgtr}
  S_{abc}=-S_{bac},\qquad S_{ab}{}^a=0,\qquad S_{abc}+S_{bca}+S_{cab}=0.
\end{equation}
In four dimensions, the totally antisymmetric component of torsion is
parameterized by the pseudovector field $T^{*d}:=-\frac16T_{abc}\ve^{abcd}$,
where $\ve^{abcd}$ is the totally antisymmetric tensor, $\ve_{0123}=1$.
Thus the torsion tensor has three irreducible components.

We change variables: $e_\al{}^a,\om_\al{}^{ab}$ $\mapsto$
$e_\al{}^a,T_{\al\bt\g}$. Now the quadratic torsion terms become simply the mass
terms for torsion which can be easily analyzed. The
transformation of variables restricts the possible choice of kinetic terms for
the vierbein, if we do not want it to be mixed with the torsion. The
corresponding invariant must have dimension $l^{-2}$, where $l$ is the dimension
of length. We have only five tensors of this dimensionality in our disposal
\begin{equation*}
  \widetilde R_{\al\bt\g\dl},\qquad R_{\al\bt\g\dl},\qquad\wn_\al T_{\bt\g\dl},
  \qquad\nb_\al T_{\bt\g\dl}
  ,\qquad T_{\al\bt\g}T_{\dl\e\z}.
\end{equation*}
The invariants are built by different contractions of these tensors with the
metric. We see that the only possible invariants are $\widetilde R$ and
$\wn_\al T^\al$. The last term mixes kinetic term of the vierbein with torsion
components. To simplify matters we postulate that this term must be absent in
dynamical torsion theory. Thus we are left with the unique possibility: the
Hilbert--Einstein action for the vierbein. Therefore, we further transform
variables: $e_\al{}^a$, $T_{\al\bt\g}$ $\mapsto$ $g_{\al\bt}$, $T_{\al\bt\g}$.

It is well known that metric in the Hilbert--Einstein action describes massless
gravitons $2^+$, and the full symmetry under general coordinate transformations
must be used for elimination of the unphysical degrees of freedom of the metric.
Therefore the problem is reduced to choosing the Lagrangian for the torsion
tensor, and, what is important, we cannot use general coordinate transformations
to eliminate its degrees of freedom.
%******************************************************************************
\section{(1+3)-decomposition of torsion}
%*******************************************************************************
Torsion tensor has three irreducible components with respect to
the Lorentz group: $S_{abc}$, $T_a$, and $T^*_a$. To analyse positive
definiteness of quadratic forms of velocities and masses, we further decompose
torsion components with respect to the rotational subgroup
$\MS\MO(3)\subset\MS\MO(1,3)$, because invariant quadratic forms for the Lorentz
group are not positive or negative definite.

Positive definiteness of the quadratic form of torsion momenta or, equivalently,
veloci\-ties depends only on the metric signature. Hence, we put
$g_{\al\bt}\mapsto\eta_{ab}:=\diag(+---)$, i.e.\ we shall work in Minkowskian
space. We perform the (3+1)-decomposition of all components: $(a)=(0,i)$,
$i:=1,2,3$. Space indices are denoted by Latin letters from the middle of the
alphabet $i,j,\dotsc:=1,2,3$, and summation over them is performed by the
Euclidean metric $\dl_{ij}$. For example, $T^aT_a=T^0T_0-T^iT_i$. Covector
components of torsion tensor have simple decompositions $(T_a):=(T_0,T_i)$ and
$(T^*_a):=(T^*_0,T^*_i)$.
In the massive case, components $T_0$ and $T^*_0$ describe scalar particles
$0^+$ and $0^-$, respectively, where indices denote parities. Covector
components $(T_i)$ and $(T^*_i)$ correspond to vector particles of spin $1^-$
and $1^+$. Irreducible component $S_{abc}$ describes more particles. Because of
the antisymmetry in the first two indices the following representations hold:
\begin{equation}                                                  \label{abvxfm}
  S_{ij0}=\ve_{ijl}H^l\quad\Leftrightarrow\quad H^l:=\frac12S_{ij0}\ve^{ijl};
  \qquad
  S_{ijk}=\ve_{ijl}H^l{}_k\quad\Leftrightarrow\quad
  H^l{}_k:=\frac12S_{ijk}\ve^{ijl},
\end{equation}
where $\ve_{ijk}$ is the totally antisymmetric third rank tensor,
$\ve_{123}=1$. The relation $S_{ijk}\ve^{ijk}=0$ implies the tracelessness
condition $H^i{}_i=0$.
Components $S_{0ij}$ are decomposed into symmetric traceless, antisymmetric
parts and the trace:
\begin{equation*}
  S_{0ij}=S_{0\lbrace ij\rbrace}+S_{0[ij]}+\frac13\dl_{ij}S,\qquad
  S:=S_{0i}{}^i.
\end{equation*}
Equality $S_{a0}{}^a=S_{00}{}^0+S_{i0}{}^i=0$ implies $S=0$. From the relation
$S_{ai}{}^a=S_{0i}{}^0+S_{ji}{}^j=0$ we deduce that
\begin{equation}                                                  \label{avbxgs}
  S_{0i0}=-\ve_{ijk}H^{jk}\quad\Leftrightarrow\quad H^{[mn]}
  =-\frac12 S_{0i0}\ve^{imn}.
\end{equation}
Finally, the equality $S_{0ij}+S_{ij0}+S_{j0i}=0$ yields the expression for the
antisymmetric part
\begin{equation}                                                  \label{nxcbfg}
  S_{0[ij]}=-\frac12\ve_{ijk}H^k\quad\Leftrightarrow\quad H^k=-S_{0ij}\ve^{ijk}.
\end{equation}

Thus, irreducible components of torsion with respect to the subgroup
$\MS\MO(3)\subset\MS\MO(1,3)$ are:
\begin{equation}                                                  \label{anbchg}
\begin{aligned}
  &S_{0i0}, & & (1^-)& & -\quad\text{3 components},
\\
  &S_{0\lbrace ij\rbrace},\quad S_{0i}{}^i=0, & & (2^+)&
  & -\quad\text{5 components},
\\
 & H_{\lbrace ij\rbrace},\quad H_i{}^i=0, && (2^-) && -\quad\text{5 components},
\\
  &H_i, && (1^+) & & -\quad\text{3 components},
\\
  &T_0, && (0^+) & & -\quad\text{1 component},
\\
  &T_i, && (1^-) & & -\quad\text{3 components},
\\
  &T^*_0, && (0^-) & & -\quad\text{1 component},
\\
  &T^*_i, && (1^+) & & -\quad\text{3 components},
\end{aligned}
\end{equation}
the antisymmetric parts $S_{0[ij]}$ and $H_{[ij]}$ being expressed through
independent components by equalities (\ref{nxcbfg}) and (\ref{avbxgs}).

In the massive case, all independent $\MS\MO(3)$-components of torsion are
propagated. If one or more components of torsion turn out to be massless, then
the model posses extra gauge symmetry and the number of propagated components is
reduced but positive definiteness of the velocity quadratic form does not
change. Therefore we focus our attention on the massive case.
%******************************************************************************
\section{Torsion mass terms}
%*******************************************************************************
Since $T_{abc}$ has three Lorentz-irreducible components, $S_{abc}$, $T_a$, and
$T^*_a$, then the most general mass term depends on three constants:
\begin{equation}                                                  \label{abvxfr}
  L_m:=-\bt_1S^{abc}S_{abc}-\bt_2T^aT_a-\bt_3T^{*a}T^*_a.
\end{equation}
If there are no tachyons, then this quadratic form must be negative definite.
Simple calculations yield formulae:
\begin{equation}                                                  \label{annvmj}
\begin{split}
  S^{abc}S_{abc}=&2S^{0\lbrace ij\rbrace}S_{0\lbrace ij\rbrace}
  -2H^{\lbrace ij\rbrace}H_{\lbrace ij\rbrace}-3S^{0i0}S_{0i0}+3H^iH_i,
\\
  T^aT_a=&(T_0)^2-T^iT_i,
\\
  T^{*a}T^*_a=&(T^*_0)^2-T^{*i}T^*_i.
\end{split}
\end{equation}
We see that for any choice of $\bt_{1,2,3}$ the quadratic form (\ref{abvxfr}) is
not negative definite. It means that those torsion components which produce
positive contribution must be nonpropagating and have no kinetic terms.

Suppose that $\bt_1>0$, then the negative contribution is given by
$S_{0\lbrace ij\rbrace}$ and $H_i$. Thus, if the kinetic term is correct, the
component $S_{abc}$ can describe only massive particles $2^+$ and $1^+$.
Components $H_{\lbrace ij\rbrace}$ and $S_{0i0}$ produce positive contribution,
correspond to tachyons, and must have no velocity squared terms in the
Lagrangian. Conversely, if $\bt_1<0$, then propagating modes of torsion
component $S_{abc}$ can be only $H_{\lbrace ij\rbrace}$ and
$S_{0i0}$. In this case, $S_{abc}$ describes only particles $2^-$ and $1^-$.

Vector $T_a$ and pseudovector $T^*_a$ components of torsion describe particles
of positive mass in the following cases. If $\bt_2>0$, then propagating mode
can be only $T_0$, which correspond to scalar particles $0^+$. For $\bt_2<0$
the negative contribution is given by vector component $T_i$ describing $1^-$
particles.

The similar situation happens for $T^*_a$ component. If $\bt_3>0$ or $\bt_3<0$,
then propagating particles can be only $T^*_0$ $(0^-)$ or $T^*_i$ $(1^+)$,
respectively.
%******************************************************************************
\section{Kinetic terms}
%*******************************************************************************
The kinetic term has dimensionality $l^{-4}$ and is built from tensors
\cite{Christ80}:
\begin{equation}                                                  \label{abbxvf}
\begin{aligned}
  &\nb_\al\nb_\bt R_{\g\dl\e\z}, &&\quad R_{\al\bt\g\dl}R_{\e\z\eta\theta},
  && \quad\nb_\al\nb_\bt\nb_\g T_{\dl\e\z},
  && \quad\nb_\al T_{\bt\g\dl}\nb_\e T_{\z\eta\theta},
\\
  & \nb_\al\nb_\bt T_{\g\dl\e}T_{\z\eta\theta},
  && \quad \nb_\al T_{\bt\g\dl}T_{\e\z\eta}T_{\theta\iota\kappa},
  && \quad T_{\al\bt\g}T_{\dl\e\z}T_{\eta\theta\iota}T_{\kappa\lm\mu},
  && \quad\nb_\al T_{\bt\g\dl}R_{\e\z\eta\theta},
\\
  & T_{\al\bt\g}\nb_\dl R_{\e\z\eta\theta},
  && \quad T_{\al\bt\g}T_{\dl\e\z}R_{\eta\theta\iota\kappa}, && &&
\end{aligned}
\end{equation}
There are 151 invariants listed in \cite{Christ80} which are too many for
analytical analysis. Neverthe\-less, note the following. The velocities squared
terms $\dot T^2$, where the dot denote time derivatives $\pl/\pl x^0$ and tensor
indices are dropped, are contained in invariants $R^2$, $\nb T\nb T$, and
$\nb TR$, and also in invariants $\nb\nb T\,T$ and $\nb T R$ after integration
by parts. In the linear approximation, all invariants quadratic in velocities
have the same form as invariants $\nb T\nb T$. Therefore, it is sufficient to
consider only invariants constructed from the tensor
$\nb_\al T_{\bt\g\dl}\nb_\e T_{\z\eta\theta}$ by different contractions with the
metric.

We start with the component $S_{abc}$. Because this component satisfies Eqs.\
(\ref{ehdgtr}), there are only three independent invariants with respect to the
full Lorentz group including reflections which are quadratic in first
derivatives. We choose them in the form
\begin{equation}                                                  \label{avcxde}
  \nb_a S_{bcd}\nb^a S^{bcd},\qquad\nb_a S_b{}^a{}_c\nb_d S^{bdc}\qquad
  \nb_a S_b{}^a{}_c\nb_d S^{cdb}.
\end{equation}
The remaining invariants can be expressed through them. The proof of this
statement is by item-by-item examination using the following properties. Two
integration by parts imply relations:
\begin{equation*}
\begin{split}
  \nb_a S_b{}^a{}_c\nb_d S^{bdc}\overset{\div}=&\nb_a S_{bcd}\nb^c S^{bad},
\\
  \nb_a S_b{}^a{}_c\nb_d S^{cdb}\overset{\div}=&\nb_a S_{bcd}\nb^c S^{dab},
\\
  \nb_a S_{bc}{}^a\nb_d S^{bcd}\overset{\div}=&\nb_a S_{bcd}\nb^d S^{bca}.
\end{split}
\end{equation*}
Moreover, Eq.\ (\ref{ehdgtr}) implies identities:
\begin{equation*}
\begin{split}
  \nb_a S_{bcd}\nb^a S^{cdb}=&-\frac12\nb_a S_{bcd}\nb^a S^{bcd},
\\
  \nb_{[a}S_{bc]d}\nb^{[a}S^{bc]d}=&\frac13\nb_a S_{bcd}\big(\nb^a S^{bcd}
  +2\nb^c S^{abd}\big).
\end{split}
\end{equation*}
Finally, there is the relation
\begin{multline*}
  0=\nb_a S_{bcd}\big(\nb^b S^{cda}+\nb^c S^{dba}+\nb^d S^{bca}\big)
  =\nb_a S_{bcd}\big(\nb^d S^{bca}+2\nb^c S^{dba}\big)=
\\
  =\nb_a S_{bcd}\big(\nb^d S^{bca}-2\nb^c S^{bad}+2\nb^c S^{dab}\big),
\end{multline*}
where Eqs.\ (\ref{ehdgtr}) are used. The last equality after integration by
parts is used to express the invariant $\nb_a S_{bc}{}^a\nb_d S^{bcd}$ through
invariants (\ref{avcxde}).

The absence of ghosts is defined by the velocity quadratic form. Therefore, we
express invariants (\ref{avcxde}) through $(1+3)$-components and, for
simplicity, keep only quadratic terms:
\begin{equation}                                                  \label{anncvb}
\begin{split}
  \nb_a S_{bcd}\nb^a S^{bcd}\quad\ni\quad&2\dot S^{0\lbrace ij\rbrace}
  \dot S_{0\lbrace ij\rbrace}-2\dot H^{\lbrace ij\rbrace}
  \dot H_{\lbrace ij\rbrace}-3\dot S^{0i0}\dot S_{0i0}+3\dot H^i\dot H_i,
\\
  \nb_a S_b{}^a{}_c\nb_d S^{bdc}\quad\ni\quad&\dot S^{0\lbrace ij\rbrace}
  \dot S_{0\lbrace ij\rbrace}-\dot S^{0i0}\dot S_{0i0}+\frac12\dot H^i\dot H_i,
\\
  \nb_a S_b{}^a{}_c\nb_d S^{cdb}\quad\ni\quad&\dot S^{0\lbrace ij\rbrace}
  \dot S_{0\lbrace ij\rbrace}-\frac12\dot H^i\dot H_i.
\end{split}
\end{equation}

The kinetic part of the Lagrangian for torsion component $S_{abc}$ has the form
\begin{equation}                                                  \label{abncbg}
  L_\Sk:=c_1\nb_a S_{bcd}\nb^a S^{bcd}+c_2\nb_a S_b{}^a{}_c\nb_d S^{bdc}
  +c_3\nb_a S_b{}^a{}_c\nb_d S^{cdb},
\end{equation}
where $c_{1,2,3}$ are coupling constants. In terms of $(1+3)$-decomposition this
Lagrangian yields the quadratic form
\begin{multline}                                                  \label{anwhyt}
  L^{(2)}_\Sk=\dot S^{0\lbrace ij\rbrace}\dot S_{0\lbrace ij\rbrace}
  (2c_1+c_2+c_3)-2\dot H^{\lbrace ij\rbrace}\dot H_{\lbrace ij\rbrace}c_1-
\\
  -\dot S^{0i0}\dot S_{0i0}(3c_1+c_2)
  +\dot H^i\dot H_i\left(3c_1+\frac12c_2-\frac12c_3\right).
\end{multline}
The analysis of the mass term for $S_{abc}$ shows that propagating can be either
separate torsion $\MS\MO(3)$-components or pairs ($S_{0\lbrace ij\rbrace}$,
$H_i$) or ($H_{\lbrace ij\rbrace}$, $S_{0i0}$).

Since we have only three invariants (\ref{avcxde}), and there are four
independent $\MS\MO(3)$-components of torsion $S_{abc}$, then the absence of
three kinetic terms implies the absence of the fourth. Therefore Lagrangian
(\ref{abncbg}) cannot describe propagation of only one
$\MS\MO(3)$-irreducible component of $S_{abc}$.

Now we consider the possibility of simultaneous propagation of two
$\MS\MO(3)$-irreducible components of $S_{abc}$. If the only propagating
components are $S_{0\lbrace ij\rbrace}$ and $H_i$, then the following equalities
must hold
\begin{equation*}
  c_1=0,\qquad c_2=0,
\end{equation*}
the constant $c_3$ being arbitrary. Then the kinetic part of the Lagrangian is
\begin{equation*}
  L^{(2)}_\Sk=c_3\dot S^{0\lbrace ij\rbrace}\dot S_{0\lbrace ij\rbrace}
  -\frac12c_3\dot H^i\dot H_i.
\end{equation*}
It is clear that for any $c_3\ne0$ the model contains ghost. Consequently,
propagation of only $S_{0\lbrace ij\rbrace}$ and $H_i$ is impossible.

Similar situation happens with components $H_{\lbrace ij\rbrace}$, $S_{0i0}$.
If they are the only propagating ones, then the equalities
\begin{equation*}
  c_3=2c_1,\qquad c_2=-4c_1
\end{equation*}
must hold, where $c_1$ is arbitrary. The corresponding kinetic part of the
Lagrangian is
\begin{equation*}
  L^{(2)}_\Sk=-2c_1\dot H^{\lbrace ij\rbrace}\dot H_{\lbrace ij\rbrace}
  +c_1\dot S^{0i0}\dot S_{0i0}.
\end{equation*}
We see that the model contains ghost for any $c_1\ne0$. Therefore Lagrangian
$L^{(2)}_\Sk$ cannot describe propagation of only one pair of components
($H_{\lbrace ij\rbrace}$, $S_{0i0}$).

Thus, there is no Lagrangian in the considered class which propagates the
component $S_{abc}$ without ghosts and tachyons.

Now we discuss the irreducible component $T_a$. It is the usual vector field.
There are two possibilities of models without ghosts and tachyons. Lagrangians
\begin{align}                                                     \label{abncgd}
  L_1:=&-\frac14\g_1F^{ab}F_{ab}+\frac12m^2_1T^aT_a,
\\                                                                \label{anfmyh}
  L_2:=&~~\frac12\g_2(\nb_aT^a)^2-\frac12m^2_2T^aT_a,
\end{align}
where $F_{ab}:=\pl_a T_b-\pl_b T_a$, $\g_{1,2}>0$ and $m_{1,2}>0$ are the only
ones without ghosts and tachyons. Lagrangian (\ref{abncgd}) describes
propagation of vector particles $1^-$ of mass $m_1$. It is the Proca field. The
second Lagrangian (\ref{anfmyh}) propagates scalar particles $0^+$ of mass
$m_2$.

The same happens for pseudovector torsion component $T^*_a$. There are only
two ghost and tachyon free Lagrangians:
\begin{align}                                                     \label{abncgo}
  L_3:=&-\frac14\g_3G^{ab}G_{ab}+\frac12m^2_3T^{*a}T^*_a,
\\                                                                \label{anfmyo}
  L_4:=&~~\frac12\g_4(\nb_aT^{*a})^2-\frac12m^2_4T^{*a}T^*_a,
\end{align}
where $G_{ab}:=\pl_a T^*_b-\pl_b T^*_a$, $\g_{3,4}>0$ and $m_{3,4}>0$.
Lagrangians (\ref{abncgo}) and (\ref{anfmyo}) do not contain ghosts and tachyons
and propagate particles $1^+$ and $0^-$ of masses $m_3$ and $m_4$, respectively.

Returning to the gravity models with dynamical models, we summarize. We have
found four Lagrangians
\begin{align}                                                     \label{abncfr}
  \frac1\vol L_{13}:=&\kappa\widetilde R(g)-\frac14\g_1F^{\al\bt}F_{\al\bt}
  +\frac12m^2_1T^\al T_\al-\frac14\g_3G^{\al\bt}G_{\al\bt}
  +\frac12m^2_3T^{*\al} T^*_\al,
\\                                                                \label{anmrty}
  \frac1\vol L_{23}:=&\kappa\widetilde R(g)+\frac12\g_2(\nb_\al T^\al)^2
  -\frac12m^2_2T^\al T_\al-\frac14\g_3G^{\al\bt}G_{\al\bt}
  +\frac12m^2_3T^{*\al} T^*_\al,
\\                                                                \label{efgsty}
  \frac1\vol L_{14}:=&\kappa\widetilde R(g)-\frac14\g_1F^{\al\bt}F_{\al\bt}
  +\frac12m^2_1T^\al T_\al+\frac12\g_4(\nb_\al T^{*\al})^2
  +\frac12m^2_4T^{*\al} T^*_\al,
\\                                                                \label{egdrwg}
  \frac1\vol L_{24}:=&\kappa\widetilde R(g)+\frac12\g_2(\nb_\al T^\al)^2
  -\frac12m^2_2T^\al T_\al+\frac12\g_4(\nb_\al T^{*\al})^2
  +\frac12m^2_4T^{*\al} T^*_\al.
\end{align}
For $\kappa>0$, $\g_{1,2,3,4}>0$, $m_{1,2,3,4}>0$ these Lagrangians do not
contain ghosts and tachyons. In all four cases, the first term is the usual
Hilbert--Einstein Lagrangian describing the massless graviton $2^+$. Lagrangians
$L_{13}$, $L_{23}$, $L_{14}$, and $L_{24}$ propagate in addition massive
degrees of freedom: $(1^-,1^+)$, $(0^+,1^+)$, $(1^-,0^-)$, and $(0^+,0^-)$,
respectively.

Mass term in the obtained Lagrangians can be set to zero. Then torsion describes
massless particles of the same spins and parities. The extra gauge symmetries
appear in these cases, and the number of propagating degrees of freedom reduces.
For example, let us put $m_1=0$ in $L_{13}$ and $L_{14}$. Then these Lagrangians
are invariant with respect to gauge transformations
\begin{equation*}
  T_\al\mapsto T_\al+\pl_\al\vf,
\end{equation*}
where $\vf(x)$ is an arbitrary local transformation parameter. These
transformations coincide with the gauge transformations in electrodynamics.
Therefore, if we switch on the interac\-tion with matter fields in the same way,
then the torsion trace $T_a$ can be identified with the electromagnetic
potential. Unfortunately, the geometric meaning of such identification is not
clear and requires further analysis.

If $m_2=0$ in Lagrangians $L_{23}$ and $L_{24}$, then they are invariant with
respect to gauge transformations
\begin{equation*}
  T^\al\mapsto T^\al+\wn_\bt\om^{\bt\al},
\end{equation*}
where $\om^{\bt\al}(x)=-\om^{\al\bt}(x)$ are local parameters. Note that the
number of independent parameters is
\begin{equation*}
  C^2_4-C^3_4+C^4_4=6-4+1=3,
\end{equation*}
which is necessary for elimination of three unphysical degrees of freedom of
vector field $T^\al$.

The same happens with the pseudovector field $T^*_a$.
%******************************************************************************
\section{Conclusion}
%*******************************************************************************
Thus, supposing that kinetic terms of vierbein $e_\al{}^a$ and torsion
$T_{\al\bt\g}$ do not mix, we proved that there are only four Lagrangians
$L_{13}$, $L_{23}$, $L_{14}$, and $L_{24}$ without ghosts and tachyons. In all
cases, the vierbein describes one massless graviton $2^+$. In addition, torsion
describes massive or massless particles $1^-$, $1^+$ ($L_{13}$); $0^+$, $1^+$
($L_{23}$); $1^-$, $0^-$ ($L_{14}$); or $0^+$, $0^-$ ($L_{24}$). Moreover, we
proved that any other Lagrangian for torsion free of ghosts and tachyons must
have the same linear approximation as one of the listed.

We proved also that the irreducible component $S_{abc}$ of torsion cannot have a
Lagrangian without ghosts and tachyons. The same result was obtained in
\cite{FabTec19} using different causal analysis. In addition, the causal
analysis yields the argument in favor of Lagrangian (\ref{abncfr}) of the
present paper \cite{Fabbri15}.

This work was performed at the Steklov International Mathematical Center and
supported by the Ministry of Science and Higher Education of the Russian
Federation (agreement no. 075-15-2019-1614).

%\bibliography{book,gravity,math,my,qft}
%\bibliographystyle{unsrt}
\end{document}